\documentclass[graybox]{svmult}

\usepackage{natbib}

\usepackage{mathptmx}       
\usepackage{helvet}         
\usepackage{courier}        
\usepackage{type1cm}        
\usepackage{makeidx}         
\usepackage{graphicx}        
\usepackage{multicol}        
\usepackage[bottom]{footmisc}

\makeindex             

\begin{document}

\title*{Outer Regions of the Milky Way}
\author{Francesca Figueras}
\institute{Francesca Figueras \at Institute of Cosmos Science (IEEC-UB), University of Barcelona, Spain \email{cesca@fqa.ub.edu}}
%
%
\maketitle

\abstract{With the start of the {\it Gaia} era, the time has come to address the major challenge of deriving the star formation history and evolution of the disk of our Milky Way.  Here we review our present knowledge of the outer regions of the Milky Way disk population. Its stellar content, its structure and its dynamical and chemical evolution are summarized, focussing  on our lack of understanding both from an observational and a theoretical viewpoint. We  describe the unprecedented data that {\it Gaia} and the upcoming ground-based spectroscopic surveys will provide in the next decade.  More in detail, we quantify the expect accuracy in position, velocity and astrophysical parameters of some of the key tracers of the stellar populations in the outer Galactic disk. Some insights on the future capability of these surveys to answer crucial and fundamental issues are discussed, such as the mechanisms driving  the spiral arms and the warp formation. Our Galaxy, the Milky Way, is our cosmological laboratory for understanding the process of formation and evolution of disk galaxies. What we learn in the next decades will be naturally transferred to the extragalactic domain.}

\section{Introduction}
\label{sec:1}  

The kinematical and chemical characterization of the stellar populations in the outer regions of the galactic disks are a crucial and key element to understand the process of disk formation and evolution. These outer regions are areas in the low-density regime and thus hard to observe in external galaxies. It is in this context that our Galaxy, the Milky Way, can truly be a  laboratory and an exceptional environment to undertake such studies. The disk formation in our Milky Way was an extended process which started about 10\,Gyr ago and continues to the present. Throughout this evolution, time-dependent dynamical agents  such as radial migration or resonant scattering by transient or long-lived structures have been driving the orbital motion of the stars. Concerning star formation and evolution, key factors such as the initial mass function and the star formation history are fundamental ingredients to describe the growth of disks. Furthermore, it has to be kept in mind that the process of disk formation also involves agents which are not yet well understood, from the dynamical influence of the puzzling three-dimensional structure of the dark matter halo to the characterization of the infalling gas which gradually builds up the disk and forms stars quiescently. The scenario becomes even more complex when other decisive factors such as mergers or gravitational interactions with satellites come into play.

Two approaches has usually been considered. From the extragalactic point of view, disks at different redshifts can be studied. This approach is limited to global information integrated over the disk stellar populations but has the advantage of tracing the evolution of disk properties with time. In a second approach, normally referred to as galactic archaeology (the approach used for the Milky Way), the disk evolution is reconstructed by  resolving the stellar populations into individual stars. Disk evolution is fossilized in the orbital distribution of stars, their chemical composition and their ages. A drawback of this approach is that this information may be diluted through dynamical evolution and radial mixing. Tracers can be stars, open clusters and/or gas. In this Chapter we will concentrate on the stellar component. Molecular gas is  described by Watson and Koda (this volume). Nonetheless, when studying the Milky Way, we should not forget the perspective that in several aspects our Galaxy is not a typical late-type spiral galaxy. The  unusually quiescent merger history of the Milky Way has been discussed in detail by \citet{vander2011}, who remark \begin{quotation} ... the unique possibility to make very detailed chemical studies of stars in the Milky Way provides an independent opportunity to evaluate the merger history of our large disk galaxy... \end{quotation}  

In Sect.~\ref{sec:stellar} we describe the current knowledge of the stellar content in the outer regions of the Galactic disk. Sect.~\ref{nonaxi} deals with the non-axisymmetric structures that most contribute to the dynamics of these regions. Later on, in Sect.~\ref{chemo}, we introduce some of the essential pieces towards a future chemo-dynamical model of the Milky Way. Finally, in Sect.~\ref{LargeSurvey}, we present a brief overview of the upcoming new astrometric and spectroscopic data both, from the {\it Gaia} space astrometry mission and from on-going and future ground-based spectroscopic surveys. 

\section{The Outer Disk of the Milky Way: Stellar Content}
\label{sec:stellar}

\begin{figure}[t]
\sidecaption[t]
\includegraphics[width=\textwidth]{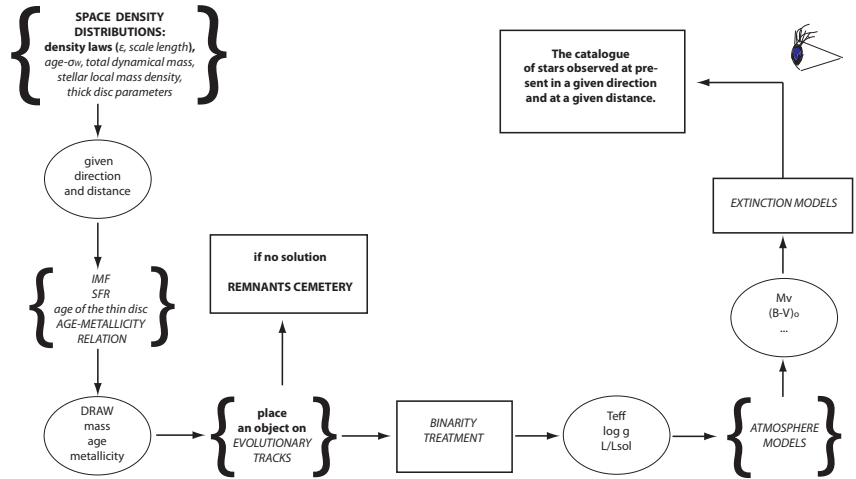}
\caption{General scheme describing the Besan\c con Galaxy Model ingredients (\citealt{czekaj2014})}
\label{fig:bgm}  
\end{figure}

In this Section we first describe some of the best tracers of the structure and kinematics of the outer regions of the Milky Way. Very useful tools such as the Besan\c con stellar population synthesis model\index{stellar population synthesis models} (\citealt{robin2014}; \citealt{czekaj2014}) allow us to quantify the number of field stars belonging to each tracer population (thin and thick disk or halo spheroid in the case of the outer regions). The key ingredients used in the strategy to generate simulated samples are shown in Fig.~\ref{fig:bgm} (from \citealt{czekaj2014}). Although the model is continuously being updated, caution has to be taken when analyzing the properties of the generated samples. The model includes several critical assumptions such us the radial scale length or the disk cut-off, which can induce significant discrepancies between model and observations. A separate paragraph is devoted to Cepheid variables, whose high intrinsic brightness  make them excellent tracers of the radial metallicity gradient (\citealt{lemasle2013}). Properties of other very good tracers such as open clusters are not treated here. The reader can find an updated characterization of this population in recent papers such as the one by \citet{netopil2016}. A summarizing discussion on the present knowledge of the outer reaches is presented in Sect.~\ref{monoceros}.

\subsection{Resolved Stellar Populations}
\label{pop-res}

\begin{figure}[t]
\begin{center}
\includegraphics[width=\textwidth]{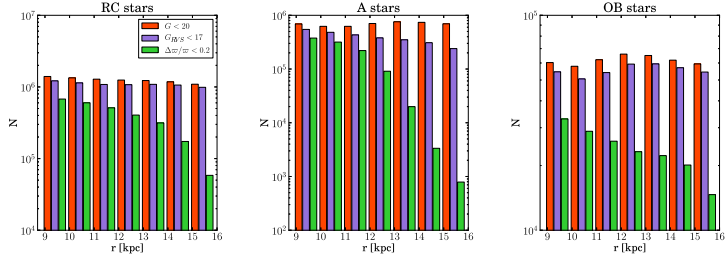}
\end{center}
\caption{Histograms of the numbers of stars in Galactocentric radius bins of 1\,kpc to be observed by the {\it Gaia} satellite (see \citealt{abedi2014} for details). The samples with {\it Gaia} magnitudes $G \le 20$, $G_{\rm RVS}\le 17$ and those with expected parallax accuracy less than 20\% (end of mission) are  shown respectively in red, purple and green. The histograms are plotted for Red Clump stars ({\it left} panel), A stars ({\it middle}) and
OB stars ({\it right})}
\label{fig:hiswarp}  
\end{figure}

\runinhead{Stellar Tracers.}  
As an example, in Fig.~\ref{fig:hiswarp} we show the  radial galactocentric distribution of three different stellar populations: Red Clump K-giant stars, and
main-sequence A and OB type stars. The estimated number of stars that {\it Gaia} will observe in the outer Milky Way region, i.e., at galactocentric radii between 9 and 16\,kpc, is shown there. These estimates come from the work of  \citet{abedi2014} where test particles were generated fitting the stellar density at the position of the Sun, as provided by the Besan\c con galaxy model (\citealt{czekaj2014}). In this case, the three-dimensional Galactic extinction model by \citet{drimmel2003} has been used. From this model we estimate that the visual extinction $A_{V}$, in the mean does not exceed $\sim$2\,magnitudes when looking toward the 
Galactic anticentre at low galactic latitudes. These low values for the interstellar extinction and the large number of stars to be fully characterized by {\it Gaia} and the future spectroscopic surveys (see Sect.~\ref{LargeSurvey}) encourage the studies of resolved stellar populations toward the Galactic anticentre and the outer regions of the Milky Way.  
 


\runinhead{Classical and Type~II Cepheids.} These pulsating intrinsically bright variable stars are excellent tracers of the extent of the thin/thick disk surface densities and also of the abundance distribution of numerous chemical elements. Whereas classical Cepheids\index{Cepheids, classical} are young stars ($< 200$\,Myr; \citealt{bono2005} associated with  young stellar clusters and OB associations, Type~II Cepheids\index{Cepheids, Type~II}, with some characteristics similar to the classical ones (e.g., period and light curve), are fainter and much older, although their evolutionary status is still not firmly established. Both are interesting targets for the outer disk. Classical and Type~II Cepheids can probably be associated with the thin and thick disk populations, respectively. Cepheids have been observed at radial galactocentric distances up to $R=18$\,kpc. Furthermore, as reported by \citet{feast2014}, a few Classical Cepheid stars have been found at approximately $1-2$\,kpc above the plane in the direction of the Galactic bulge, at distances $13-22$\,kpc from the Galactic centre. Their presence, far from the plane, suggests that they are in the flattened outer disk, and thus are excellent tracers of this at present very unknown structure.

\subsection{The Outer Reaches}
\label{monoceros}

\runinhead{The disk cut-off.} \citet{vander1981}  reported an  apparent and sudden  drop\index{disks, cut-off}  in  the  surface brightness of several edge-on galaxies at a radius of about four disk scale-lengths. This has been a long standing question both in external galaxies and in our Milky Way. Nowadays, when looking at external galaxies, one plausible explanation is that the reported edges are, in fact, inflections in the stellar density, i.e., breaks in the exponential density profiles (\citealt{bland2016}). Does our Milky Way have such a truncation? A first analysis using deep optical star counts at a low-extinction window in the Galactic anticentre direction showed a clear signature of a sharp cut-off in the star density at about 5.5 to 6\,kpc from the Sun (\citealt{robin1992}).  Later on, \citet{momany2006}, using 2MASS data,  inferred robust evidence  that there is no radial disk truncation at $R = 14$\,kpc. More recently \citet{minniti2011}, using UKIDSS-GPS and VVV surveys, pointed out that there is an edge of the stellar disk at about $R_{\rm GC} = 13.9 \pm 0.5$\,kpc along various lines of sight across the galaxy. When analyzing these data it has to be kept in mind that changes in the star counts induced by the warp and flare may not be negligible. New data seem to  disagree with a sharply truncated nature, and proof of this is the presence of stars and even star formation regions beyond the break radius. \citet{lopez2014}, using also star count techniques and Sloan Digitized Sky Survey (SDSS) data, quantified the change of the vertical scale height  with galactocentric radius of the Galactic thin and thick disks. The presence of the Galactic flare, quite prominent at large $R$,  can explain the apparent depletion of in-plane stars that is often confused with a cut-off at $R \sim 14-15$\,kpc. Furthermore, in a recent paper, \citet{carraro2016} studied the spatial distribution of early-type field stars an open clusters in the third Galactic quadrant of the Milky Way. Their Fig. 12 summarizes the spatial distribution of the young population. According to these authors the field star sample extends  up to 20\,kpc from the Galactic centre, with no indication of the disk cut-off truncation at 14\,kpc from the Galactic centre  previously postulated by \citet{robin1992}. 

\begin{figure}[t]
\includegraphics[scale=.42]{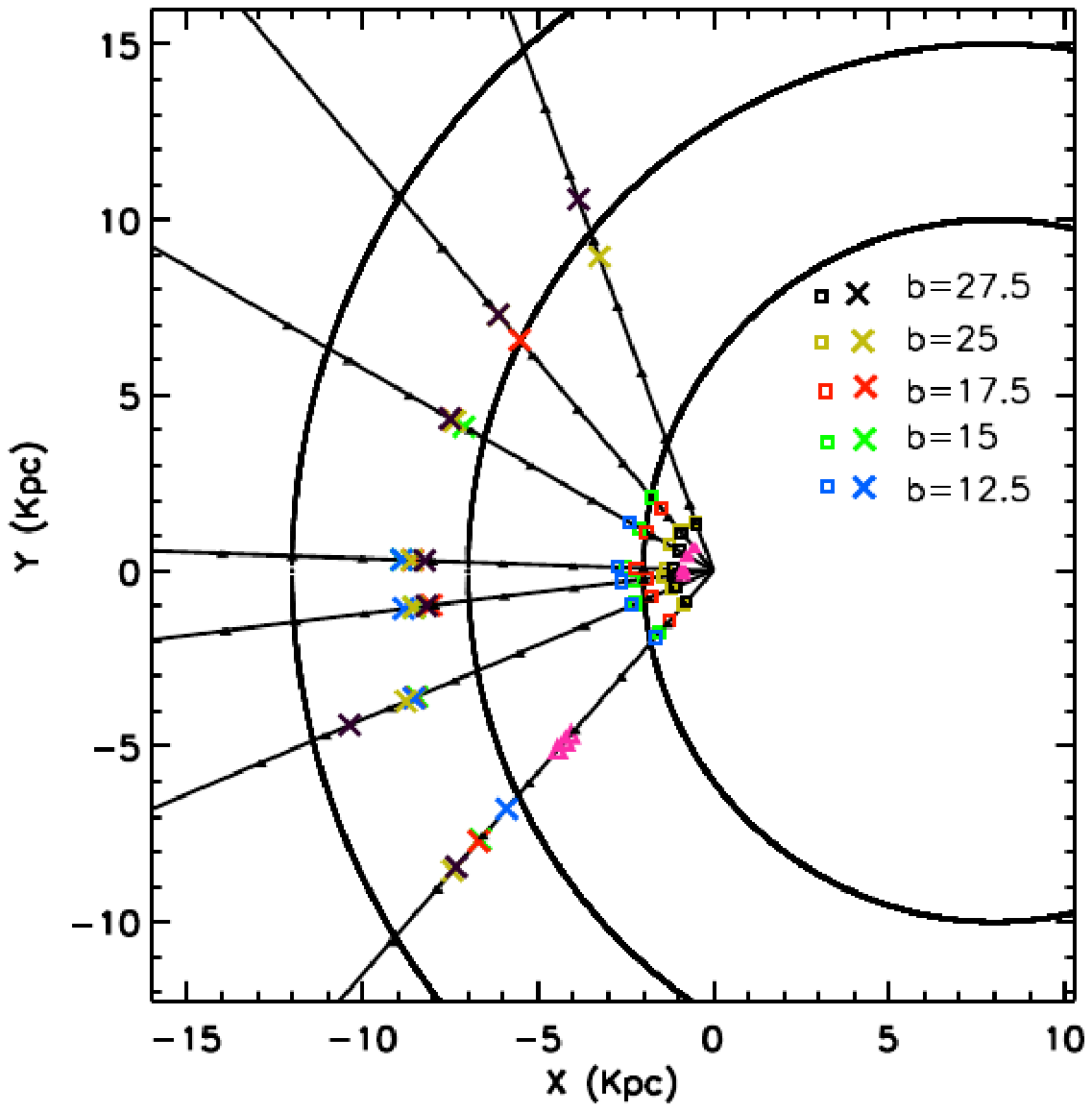}
\includegraphics[scale=.41]{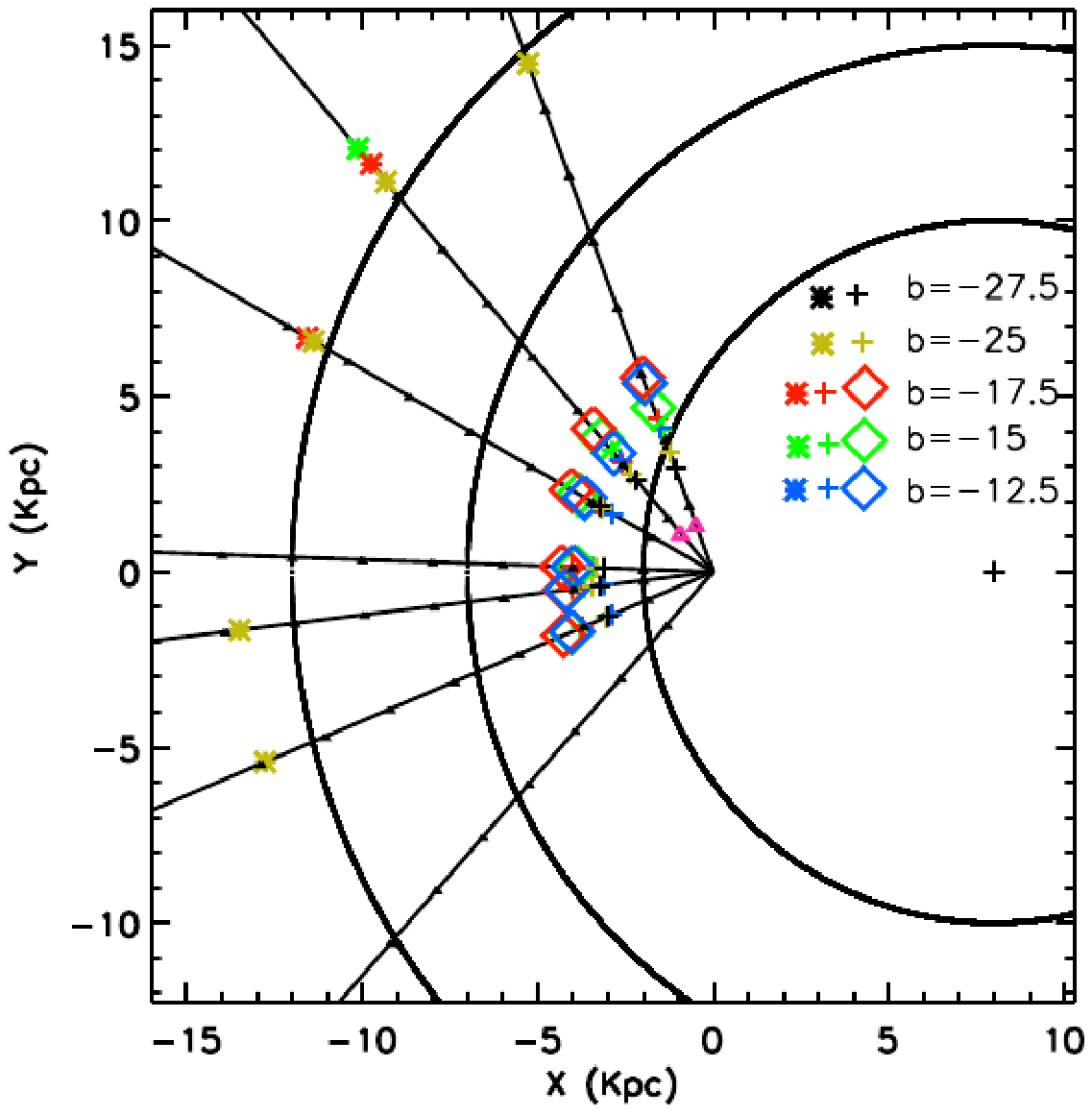}
\caption{Stellar overdensities above ({\it left}) and below ({\it right}) the plane reported by \citet{xu2015}, using SDSS data. The $ (X,Y)$ coordinates are centred at the position of the Sun}
\label{fig:outer}  
\end{figure}

\runinhead{The Monoceros Ring and beyond} The Monoceros Ring\index{Monoceros Ring}, a coherent ringlike structure at low Galactic latitude spanning about $100\,\deg$ and discovered  by \citet{newberg2002} thanks to the SDSS, deserves special attention.  This structure was first identified as an overdensity of stars at $\sim10$\, kpc, with a metallicity in the range $-1 < {\rm [Fe/H]} < 0$ (with a substantial scatter). It is a poorly understood phenomenon with no clear association to other structures such as the Canis Major or the Triangulum-Andromeda overdensity. Recent observations from PAN-STARSS1 (\citealt{slater2014})  indicate a larger extent of the stellar overdensity, up to $|b| \sim 25-35^{\circ}$ and for about $130^{\circ}$ in galactic longitude. Its origin and gravitational interaction with the Milky Way are unclear (see Sect.~\ref{nonaxi-monoceros}). More recently, \citet{xu2015}, using SDSS data,  have reported the existence of an oscillating asymmetry in the main-sequence star counts on either side of the Galactic plane in the anticentre region. These stellar overdensities, identified in a large Galactic longitude range $[110^{\circ},229^{\circ}]$, are oscillating above and below the plane and  have been observed up to distances of about $12-16$\,kpc from the Sun. As shown in Fig.~\ref{fig:outer}, the three more distant asymmetries seem to be roughly concentric rings, open in the direction of the Milky Way spiral arms. The Monoceros ring is identified as the northernmost of these structures and the others are the so-called Triangulum Andromeda overdensities (first detected by  \citealt{rocha2003}),  which could extend up to at least  25\,kpc from the Galactic centre.

\section{The Milky Way Outer Disk: Structure and Dynamics}
\label{nonaxi}

Here we will focus on those non-asymmetric structures that most contribute to the dynamics of the outer regions of the Milky Way. For an exhaustive and updated description of other components such as the Galactic bar or inner spirals, the reader is referred to the recent review by \citet{bland2016}.

\subsection{Spiral Arm Impact on Disk Dynamics and Structure}
\label{nonaxi-arms}

As is well known, spiral arms\index{spiral arms, structure} have a great impact on the evolution of galactic disks, from driving the formation of massive clouds to the perturbation of the stellar orbits of the old populations. Phenomena such as resonant trapping at corotation and Lindblad resonances, streaming motion, and radial migration (\citealt{sellwood2002})  can be understood in this context. Furthermore, the impact of the evolution of these non-axisymmetric structures on the current radial chemical gradient of the Galactic disk or on the age-metallicity relation, cannot be understood without considering spiral arms (see Sect.~\ref{chemo}). 

Accurate kinematic data are, without doubt, a first requirement to distinguish between spiral structure theories and thus on the  mechanisms of formation and evolution of spiral arms. Some of the current theories under investigation are:  the density wave theory  (\citealt{lin1964}), with a rigid spiral density wave travelling through the disk;  swing amplification (\citealt{toomre1981}; \citealt{masset1997}), with perturbations on the disk that can be swing-amplified;  invariant manifolds (\citealt{romero2006}; \citealt{athanassoula2012}), with manifolds originated in the periodic orbits around the equilibrium points;  external interactions (\citealt{sellwood1984}), for a certain fraction of mass accretion; or  chaotic orbits (\citealt{voglis2006}; \citealt{patsis2006}), important for a large perturbation, especially near corotation. It is difficult to test these theories with current available data and, complicating the situation even more, several of them may coexist. As an example, the invariant manifold mechanism has been tested using  $N$-body  simulations of  barred galaxies (\citealt{athanassoula2012}; \citealt{roca2013}). A detailed characterization of the motion of the particles through the arms or crossing them is required (\citealt{antoja2016}).  It is clear that an accurate knowledge of the kinematics is mandatory to distinguish between theories (see Sect.~\ref{LargeSurvey}).

Recently, \citet{monguio2015} published the first detection of the field star overdensity in the Perseus arm towards the Galactic anticentre.  Using a young population of B and A type stars, the authors placed the arm at $1.6 \pm 0.2$\,kpc from the Sun, estimating its stellar density amplitude to about  10\%. Moreover, these authors show how its location matches a variation in the dust distribution congruent with a dust layer in front of the arm. This favours the  assumption that the Perseus arm is placed inside the corotation radius of the Milky Way spiral pattern. The obtained heliocentric distance of the Perseus arm is slightly smaller than the 2.0\,kpc recently proposed by \citet{reid2014}. These authors used VLBI trigonometric parallaxes and proper motions of masers, thus star-forming regions, to accurately locate many arms segments in the Galactic disk. The outer spiral arm seems to be located at 13.0 $\pm$ 0.3\,kpc from the Galactic centre, with an amplitude 0.62 $\pm$ 0.18\,kpc and a pitch angle of 13.8 $\pm$ 3.3\,deg. Another critical issue is the age dependence of the radial scale length. Several  determinations can be found in the literature. Recently, \citet{monguio2015}  derived it using the Galactic disk young population. They obtained values of 2.9 $\pm$ 0.1\,kpc for B4-A1 type stars and 3.5 $\pm$ 0.5\,kpc for B4-A0 stars. Unfortunately, the uncertainty associated with these data still prevents  a direct application of these results in discriminating between inside-out or out-inside formation scenarios.

\subsection{The Galactic Warp and Flare}
\label{nonaxi-warp}

It is widely accepted that warps\index{warps, Galactic} of disk galaxies are a common phenomenon (as common as spiral structure), yet warps are still not fully understood (see \citealt{garcia2002} for a historical review). From the time when the first 21\,cm observations of our Galaxy became available, the large-scale warp in the H{\sc i} gas disk has been apparent (\citealt{burke1957};\citealt{westerhout1957}). More than fifty years later, Levine et al. (2006) re-examined the outer H{\sc i} distribution, proposing a more complex structure with the gaseous warp well described by two Fourier modes.  The warp seems to start already within the Solar circle. \citet{reyle2009}, using 2MASS infrared data, found the stellar component to be well modelled by an S-shaped warp with a significantly smaller slope that the one seen in the H{\sc i} warp. Since then, several authors have tried to set up  the
 morphology of the Galactic warp, reaching no clear conclusion. More importantly, the current uncertainties do not allow us  to disentangle which mechanisms can explain it. Critical issues are the kinematics of the warped population and  basic properties such as its stellar age dependence.  Data available currently (PPMXL proper motions) allowed  \citet{lopez2015} to perform a  vertical motion analysis of the warp toward the Galactic anticentre. They point out  that whereas the  main S-shaped structure of the warp is a long-lived feature, the  perturbation that produces an irregularity in the southern part is most likely a transient phenomenon. Again, this is a complex kinematic feature from which a definitive scenario is difficult to  constrain. We refer to the recent work of  \citet{abedi2014} to know the  capabilities of {\it Gaia} in detecting and characterizing the kinematic properties of the warp. Accurate proper motions and radial velocities (see Sect.~\ref{LargeSurvey}) will certainly add a new dimension to this study.
 
 \begin{figure}[t]
\includegraphics[width=\textwidth]{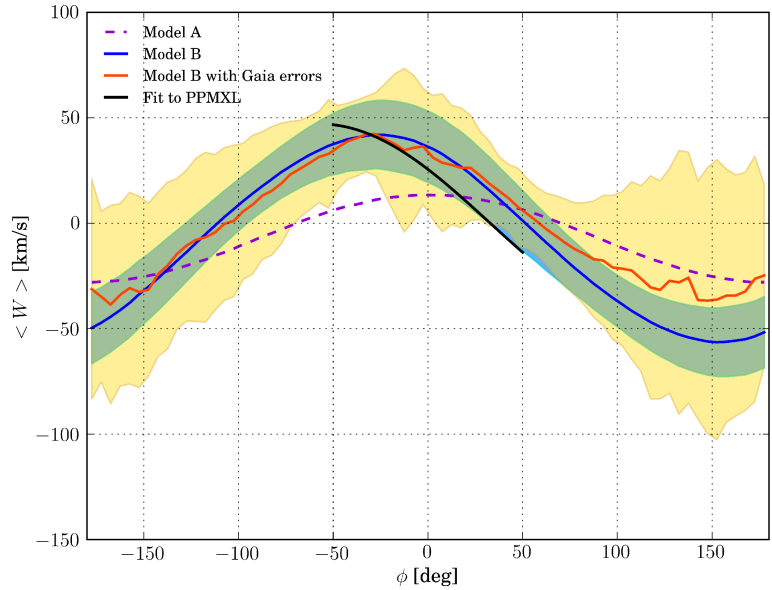}
\caption{The effect of the Galactic warp on the distribution of the mean heliocentric vertical velocity component (W) as a function of galactocentric azimuth for Red Clump stars in the galactocentric ring $13<R<14$\,kpc. Two warp models (A and B) described in \citet{lopez2015} are  plotted here (dashed purple and solid blue lines) and compared to the best fit to PPMXL proper motion data (in black). See \citet{lopez2015} for details}
\label{fig:warp}  
\end{figure}

\begin{figure}[t]
\sidecaption[t]
\includegraphics[scale=.26]{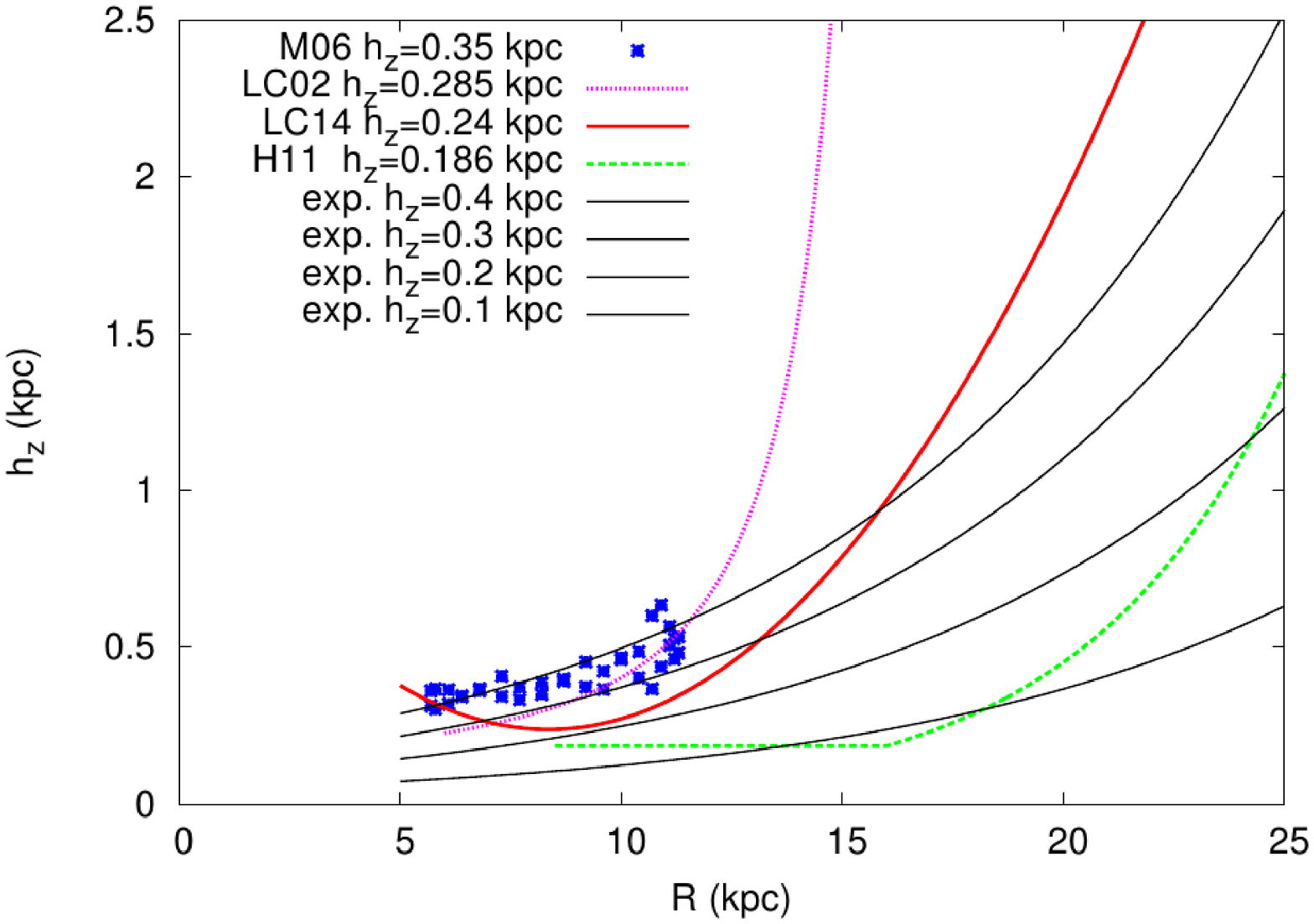}
\includegraphics[scale=.28]{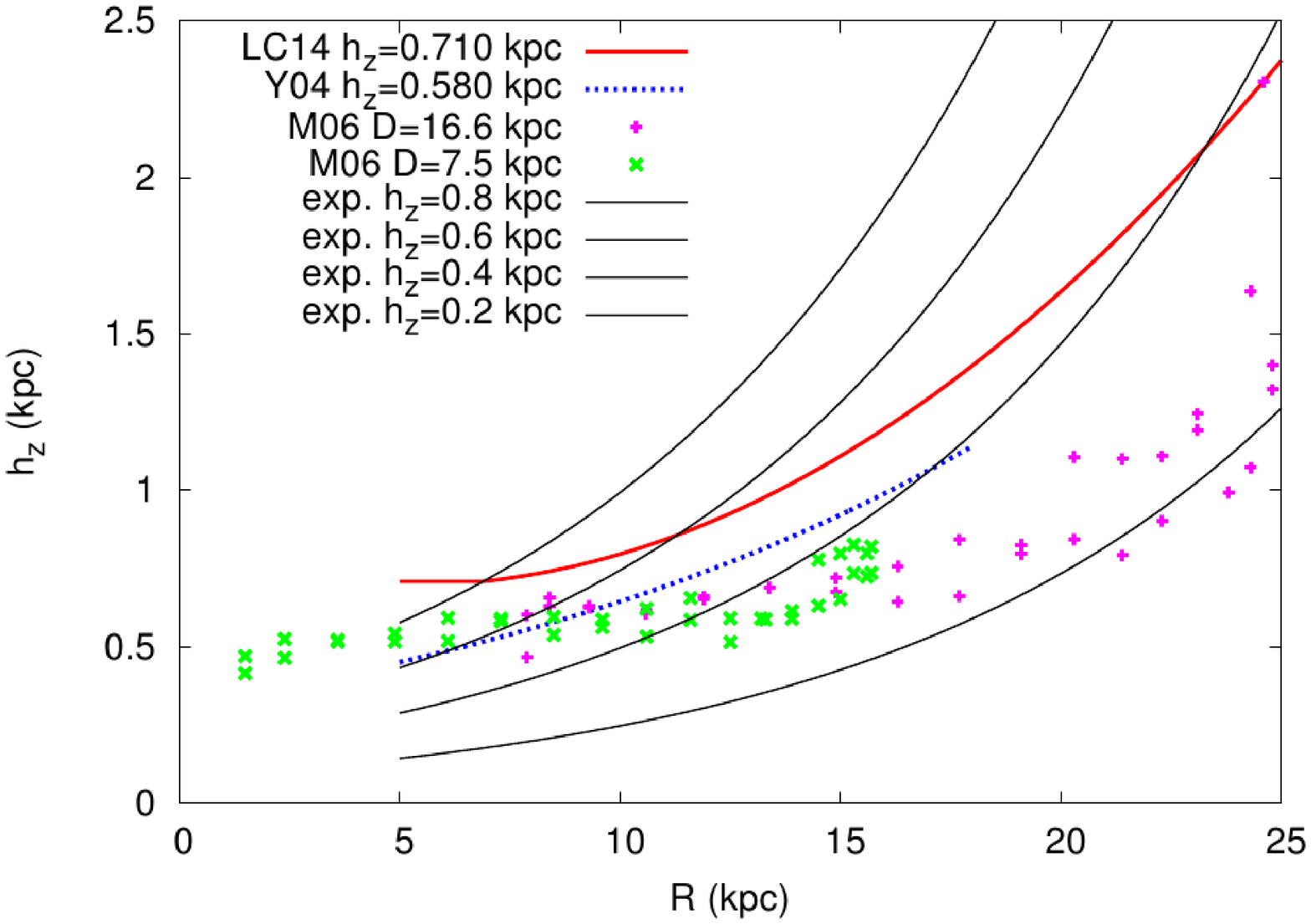}
\caption{Comparison by \citet{kalberla2014} of the current observational data (points) and exponential scale heights fits (continuous lines) to quantify the flare of the Galactic thin ({\it left}) and thick ({\it right}) components. See details in their paper}
\label{fig:flare}  
\end{figure}

\citet{lozinskaya1963} were the first to describe the observational evidence of a flare in the H{\sc i} gas in the Milky Way. Since then, the question of whether the stellar component takes part in this flaring has been a matter of debate.  \citet{momany2006} presented a first comparison of the thickness of the stellar disk, neutral hydrogen gas layer and molecular clouds. The stellar flare was traced using Red Clump and red giant stars. These authors found that the variation of the disk thickness (flaring)---treating a mixture of thin and thick stellar populations---starts at $R = 15$\,kpc and increases gradually until reaching a mean scale-height of $\sim 1.5$\,kpc at $R = 23$\,kpc. More recently,   \citet{kalberla2014} confirmed these results,  showing strong evidence for  a common flaring of gas and stars in the Milky Way. Several sources such as H{\sc i} gas,  Cepheids, 2MASS, SDSS, and Pulsar data show an increase of the scale height, growing with galactocentric radius (see Fig.~\ref{fig:flare}). Even more, authors proposed that flaring at large galactocentric distances could be stronger for the thin than for the thick disk. Although this is still a matter of debate, with {\it Gaia} we will have the opportunity not only to analyze the structure, but also the kinematic properties of these outer populations.

Several models have been proposed to account for this flare. Whereas \citet{kalberla2007} explored several mass models reflecting different dark matter distributions, others, such as \citet{minchev2012}, by preassembling $N$-body disks, showed that purely secular evolution could lead to flared disks. Others, like \citet{roskar2010}, proposed misaligned  gas  infall to provoke the flaring of the Galactic disk. Recent observational analysis of SDSS data points towards a smooth stellar distribution  (\citealt{lopez2014}), thus supporting a continuous structure for the flare and not a combination of a Galactic disk plus some component of extragalactic origin. 

\subsection{Gravitational Interaction with Satellites}
\label{nonaxi-monoceros}

Several approaches has been undertaken in the last years to characterize the effects on the Galactic disks induced by dynamical perturbations by satellites\index{satellite galaxies} (e.g., \citealt{purcell2011}; \citealt{gomez2013}; among others). All of them reinforce the picture that the Galactic disk can exhibit complex structure in response to close satellite passages, from tidal debris in a disrupted dwarf galaxy to a strong gravitational perturbation by the accretion of a satellite. Work is in progress from both the observational and the theoretical side. As an example, \citet{penarrubia2005}  proposed a model for the formation of the Monoceros ring by accreted satellite material. Their model has  recently been compared to PAN-STARSS1 data (\citealt{slater2014}) at different distances, showing broad agreement with the observed structure at mid-distances but significant differences in the far regions. These and newer simulations have since been compared to the data.  Recently, \citet{gomez2016} studied the vertical structure of a stellar disk obtained from a fully cosmological high-resolution hydrodynamical simulation of the formation of a Milky Way-like galaxy. The disk's mean vertical height can have amplitudes as large as 3\,kpc in its outer regions as a  result of a satellite-host halo-disk interaction. The simulations reproduce, qualitatively, many of the observable properties of the Monoceros Ring. Nonetheless, as pointed out by \citet{slater2014}, the crucial question for future simulations is whether such Monoceros Ring-like features can be created without causing such an unrealistically large distortion of the disk; maybe less massive satellites or particular infall trajectories could be more favourable. 

\subsection{Dynamics of the Vertical Blending and Breathing Modes}
\label{nonaxi-blend}

We currently have  observational evidence of oscillations of the Milky Way's stellar disk in the direction perpendicular to the Galactic midplane. Some of these are possibly related to the Monoceros and TriAnd overdensities in the outer disk (Sect.~\ref{monoceros}), while others are more local, within $\sim$ 2\,kpc of the Sun.  \citet{widrow2012}, using SDSS/SEGUE, RAVE, and LAMOST data, found a Galactic North-South asymmetry in the number density and bulk velocity of Solar neighbourhood stars which showed a gradual trend across the Galactic midplane and thus the appearance of a wavelike perturbation. This perturbation has the characteristics of a breathing mode, with compression and rarefaction motions, and with the displacements and peculiar velocities having opposite signs above and below the plane. \citet{widrow2014} demonstrate that both breathing and bending modes can be generated by a passing satellite or dark matter subhalo, with the nature of the perturbation being controlled by the satellite's vertical velocity relative to the disk (a slow-moving satellite would induce a bending mode whereas higher vertical velocities would induce breathing modes). More recently,  \citet{monari2016} discussed that breathing modes can be induced by several effects such as  bar or spiral perturbation,  spiral instabilities, or a possible bombarding by satellites.  Even the existence of a dark matter substructure could play a role. These authors have derived explicit expressions for the full perturbed density function of a thin disk stellar population in the presence of non-axisymetric structures such as spiral arms and bar. Undoubtedly, upcoming data will yield a more accurate and complete map of bulk motions in the stellar disk, up to about $3-4$\,kpc from the Sun (Sect.~\ref{LargeSurvey}).

\section{Towards a Chemodynamical Model of the Galactic Disk}
\label{chemo}

 \citet{prantzos2008} and \citet{prantzos2011} are two excellent and pedagogical papers  to review the basic principles and hot topics for the development of galactic chemical evolutionary models\index{chemodynamical models}. The state of the art is well summarized there. These models shall assume, among other factors, the evolution of several chemical elements up to the iron peak and the different compositions for the infalling material. Accurate observations are required to test these models, the products of which are usually expressed in terms of radial   abundance  gradients of several elements (C, N, O, Ne, Mg, Al, Si, S, Ar, Fe) and their time evolution. Thanks to the extent of present and future large surveys (see Sect.~\ref{LargeSurvey}), any formation model must be able to account not only for local, but also for radial and vertical large-scale chemical element distributions.

Radial migration has been firmly accepted to be an inseparable part of disk evolution in numerical simulations (\citealt{sellwood2002}). In radial migration models (e.g., \citealt{schonrich2009}), metal-poor stars born in the outer disk move inward to the Solar neighbourhood, while metal-rich stars born in the inner disk migrate outward. These authors suggest two mechanisms for this motion\footnote{Following the \citet{schonrich2009} terminology, churning implies a change of guiding-centre radii, while blurring means a steady increase of the oscillation amplitude around the guiding centre}: {\it blurring}, due to the scattering and subsequent increase of eccentricities over time, and {\it churning}, mostly triggered by resonant scattering at corotation. These effects would produce a large heterogeneity in the chemical abundance in the Solar neighbourhood and its environment.

\subsection{Age-Metallicity-Kinematics relations}
\label{chemo-kin}

It is well established that the velocity dispersion of the thin-disk population increases with age, a well-known phenomenon often referred to as disk heating (\citealt{binney1987}). \citet{aumer2009} used a power law to fit the thin disk  age-velocity relation (AVR) in the Solar neighbourhood,  using the most accurate data at that time: the  Hipparcos astrometry and photometric and spectroscopic data from the Geneva-Copenhagen survey (\citealt{holmberg2007}). They favour continuous heating but, as proposed also by \citet{seabroke2007},  a saturation at ages $\geq 4.5$\,Gyr could not be excluded. Concerning the age-metallicity distribution (AMD), it is important to mention the work done by \citet{haywood2006}. Biases were  evaluated in detail by this author and a new AMD with a mean increase limited to about a factor of two in $Z$ over the disk age was proposed. Again, it was emphasized that dynamical effects and complexity in the AMD clearly dominate. All the above relations have been established using data in the Solar neighbourhood. Nonetheless, and as pointed out by \citet{casagrande2011}, the Solar neighbourhood is not only assembled from local stars, but also from stars born in the inner and outer Galactic disk that migrated to their present positions (\citealt{schonrich2009}). New surveys (see Sect.~\ref{LargeSurvey}) will open a new window and different approaches to these two key and fundamental relations. 

In this direction, a new stellar chemo-kinematic relation has  recently been derived by  \citet{minchev2014} using RAVE and deeper surveys such as  SEGUE. Stars with [Mg/Fe]$ \geq 0.4$\,dex  show a peculiar kinematic behaviour. These authors used this index as a proxy of the stellar age to identify the oldest stars in the sample. These stars, born during the first year of the Galaxy's life, have  velocity dispersions  too large to be accounted for by internal disk heating.  \citet{minchev2014} showed that a chemo-dynamical model incorporating massive mergers in the early Universe and a subsequent radial migration of cool stars could explain the observed trends. More imprints such the ones reported there should be expected in the outer regions of the Galactic disk, so this work is a good example of the chemo-kinematic relations that the combination of future {\it Gaia} spectroscopic surveys will provide. They will surely bring new constraints to the formation scenarios of galactic disks. 

\subsection{The Galactic Thick Disk}
\label{chemo-thick}

The characterization of the thick disk\index{thick disks} is an important milestone when trying to understand the assembly of disk galaxies. Despite the increasing amounts of observational data in our local environment, to date we have lacked the means to discriminate among different thick disk formation scenarios for the Milky Way. A first key and basic question which arises is: do the thin and thick disks have a different origin? Two approaches are being used to answer this question, observational evidence and galaxy modelling. Observational evidence at present is,  in some sense, highly inconsistent. Whereas \citet{bovy2012a}, using SDSS data, clearly favoured a vertical structure composed of a smooth continuum of disk thicknesses with no discontinuity between the thin and thick disk, a bimodal distribution in ([Fe/H],[$\alpha$/Fe]) relation with two sequences of high- and low-$[\alpha$/Fe] seems well established (\citealt{adibekyan2012}; \citealt{nidever2014}) with high-[$\alpha$/Fe] values more prominent in the inner disk, and lower values dominating the outer parts. In this line, \citet{recio2014} point toward a clear kinematic distinction between thin and thick disk (two distinct populations). Concerning the structure, the thick disk occupies (in agreement with its hotter kinematics) a larger vertical volume around the midplane (e.g., \citealt{juric2008}), perhaps at a shorter scale-length than the thin disk (\citealt{bensby2011}; \citealt{robin2014}) and a clear  uncertainty in the radial metallicity gradient,  as will become evident in Sect.~\ref{chemo-GES}. Nonetheless, as discussed by \citet{bland2016}, one of the strongest pieces of evidence of the existence of the thick disk will be a robust statistical confirmation of its unique chemistry (see the important work by \citealt{bensby2014} in this line).

Several models has been proposed to explain the existence of the thick disk: 1) a heating process of the Galactic disk  due to satellite mergers (e.g., \citealt{abadi2003}); 2) the formation of a puffed-up structure by mere radial migration (\citealt{sellwood2002}); 3) a so-called ``upside-down'' disk formation, where old stars were formed in a relatively thick component while younger populations form in successively thinner disks (e.g., \citealt{bird2013}) or 4)  {\it in situ} formation by  early accretion of gas. The scenario for the Milky Way thick disk is presently  unclear. Several observational constraints have to be fixed. As an example, some of these models predict a vertical gradient in [$\alpha$/Fe], others do not. The {\it in situ} formation or the direct accretion of small satellites would be ruled out if $\alpha$-gradients are observed (\citealt{recio2014}). Progress in this field is further discussed in Sect.~\ref{chemo-GES}. Other future constraints could be, as pointed out by Lemasle (private communication),   the ratio of classical versus Type 2 Cepheids. The galactocentric radial change of these ratios, available from {\it Gaia}, would help us to analyze the extent of the thin/thick disk structures.

\subsection{The Radial Abundance Gradients}
\label{chemo-grad}

Examples of two reference papers on chemical evolutionary models that show different but complementary approach are those of \citet{chiappini1997} and \citet{pilkington2012a}. \citet{chiappini1997} developed a new  model for the Galaxy assuming two main infall episodes for the formation of the halo-thick disk and thin disk. The model also predicts the evolution of the gas mass, the star formation rate, the supernova rates, and the abundances of 16 chemical elements as functions of time and Galactocentric distance. In summary, a long list of detailed model results can be constrained when compared with observational data.  The approach of \citet{pilkington2012a} is different and complementary. These authors used  cosmological hydrodynamical simulations of dwarf disk galaxies to analyze the distribution of metals. Both approaches require the comparison of models and data and hence the selection of good stellar tracers, as discussed below. 

The differences between the models concern effects such us the efficiency of the enrichment processes in the inner and outer regions and the nature of the material (primordial or pre-enriched falling from the halo onto the disk). These questions can help answer fundamental questions such as whether our Galactic disk has a flattening or a steepening  radial metallicity gradient\index{abundance gradients, Galactic} with time. 
Open clusters and Cepheids are proposed as good tools to derive the time evolution of the metallicity gradients as the need to answer this question is pressing. As an example, in the future WEAVE spectroscopic survey (Sect.~\ref{LargeSurvey}) the requirement on the accuracy of the open cluster metallicity is ~0.1\,dex over the full age and metallicity range.

\begin{table}
\caption{Some of the ongoing and future ground-based surveys for chemodynamical studies of field stars in the outer regions of the Milky Way}
\label{tab:1}     
\begin{tabular}{p{1.6cm}p{2cm}p{2cm}p{1.5cm}p{4.2cm}}
\hline\noalign{\smallskip}
Survey & Dates & magnitude & Resolution & Trace populations \\
\noalign{\smallskip}\svhline\svhline\noalign{\smallskip}
{\it Gaia}-ESO & 2011-2016  & $V \sim$ 16-20 & $\sim$20000 &Red Clump giants \\
{\it Gaia}-ESO & 2011-2016  & $V<$ 16& $\sim$45000 &thick disk population \\
LAMOST   & 2012-   &  $r\sim 19-20$ & 1800 & North hemisphere \\
APOGEE & 2011-    & $H<12.2$ & 22500 & red giants and subgiants \\
EMIR  &  2017-    & $J, K \sim$ 18 & 4000 & red giants \\
WEAVE & 2018-2022 & $V=$ 16-20 & 4000 & disk OBA and Red Clump giants \\
WEAVE & 2018-2022 & $V<$ 16-17 & 20000 & Anti-centre OBA and RC disk \\
4MOST & 2022 -    &      $V\leq$ 20      &   $>$2000    &   from Sun-like to RGBs            \\
\noalign{\smallskip}\hline\noalign{\smallskip}
\end{tabular}
\end{table}

\runinhead{Tracers of chemical abundance gradients.}
As mentioned by \citet{boissier1999}, all the existing observational data only inform us on the present properties of the Galactic disk, but not on its past history, except for the tentative indications obtained from the study of abundances in planetary nebulae or in open clusters. \citet{alibes2001} reviewed the main observables for the disk of the Milky Way, which can be compared with the products of a chemical evolutionary model. As emphasized by these authors, homogeneous samples are required to derive radial gradients.  The young population (age $< $1\,Gyr) provides information on the  present-day abundances in the Galactic disk, the most important tracers being the H{\sc ii} regions and the B-type stars. H{\sc ii} regions have the drawback of large uncertainties due to the derivation of electron temperature. The B-type stars are bright, thus observed up to large distances,  with a  photospheric composition which supposedly reflects the composition of the material from which they formed, except for high rotators where there could have been some mixing of atmospheric and core material. Tracers older than 1\,Gyr provide insight into the past evolution of the abundance gradients, and thus into the evolution in abundance profiles. Among these are the planetary nebulae, with large uncertainties in distances, ages and masses; open clusters, which can be as old as 8\,Gyr; and the FGK, dwarfs and giants, a population selected as tracer in most of the ongoing large spectroscopic surveys\index{spectroscopic surveys} (see Table~\ref{tab:1}).

\subsubsection{Cepheids as Tracers of Galactic Abundance Gradients }
\label{chemo-cep}

Since the early work by \citet{harris1981}, the determinations of the Galactic [Fe/H] gradient from Galactic Cepheids\index{Cepheids, classical} have been remarkably  consistent, all of them reporting values of $\sim -0.06$\,dex/kpc (\citealt{lemasle2013}). From open clusters, a linear gradient of approximately $-0.06$\,dex/kpc has been proposed (e.g., \citealt{bragaglia2008}) but with a flattening between 10 and 14\,kpc, that is in the outer regions of the Galactic disk. This possible flattening of the abundance gradients is a long-standing question. Whereas it is not observed from Galactic Cepheids (\citealt{lemasle2013}),  a clear flat (bi-modal) gradient in the outer disk is derived from open clusters. \citet{lemasle2013} measured the Galactic abundance gradient of 11 chemical elements using high-resolution spectra of more than sixty Galactic Cepheids and focussed their analysis on the outer Galactic disk. They confirm the  existence of a gradient for all the heavy elements and noted that current data are not supporting a flattening of the gradient in the outer disk. Later on, the same team (\citealt{genovali2014}) compiled homogeneous data for a sample of 450 Cepheids deriving a linear  metallicity gradient over a broad range of Galactocentric distances ($R \sim$ 5-19\,kpc). These authors pointed out clear evidence for the increase in the intrinsic scatter when moving toward the outer disk. It cannot be ruled out that this scatter is due to  contamination by Type~II Cepheids misclassified as classical Cepheids (Sect.~\ref{monoceros}). 

\subsubsection{Recent Outcomes of the {\it Gaia}-ESO Survey}
\label{chemo-GES}
From the current surveys (SEGUE, RAVE, APOGEE, GALAH, {\it Gaia}-ESO Survey, among others), the large-scale characteristics of the thin and thick disks have started to emerge. It is difficult to summarize here the huge amount of new results published in the last $5-10$ years. In this Section, as it is impossible to make an exhaustive review, we will take one of them as a example, in this case the effort being done by the {\it Gaia} European reseach community (more than 300 scientists) to build the {\it Gaia}-ESO Survey (hereafter GES). The survey, planned for five years (2011-2016) and still ongoing,  uses the FLAMES/GIRAFFE and UVES spectrographs on the Very Large Telecope (ESO), and important preliminary results have already being published. Two approaches have been followed to derive the vertical and radial trends of the chemical gradient, from open clusters and from FGK main-sequence field stars.

\begin{figure}[t]
\sidecaption[t]
\includegraphics[scale=.14]{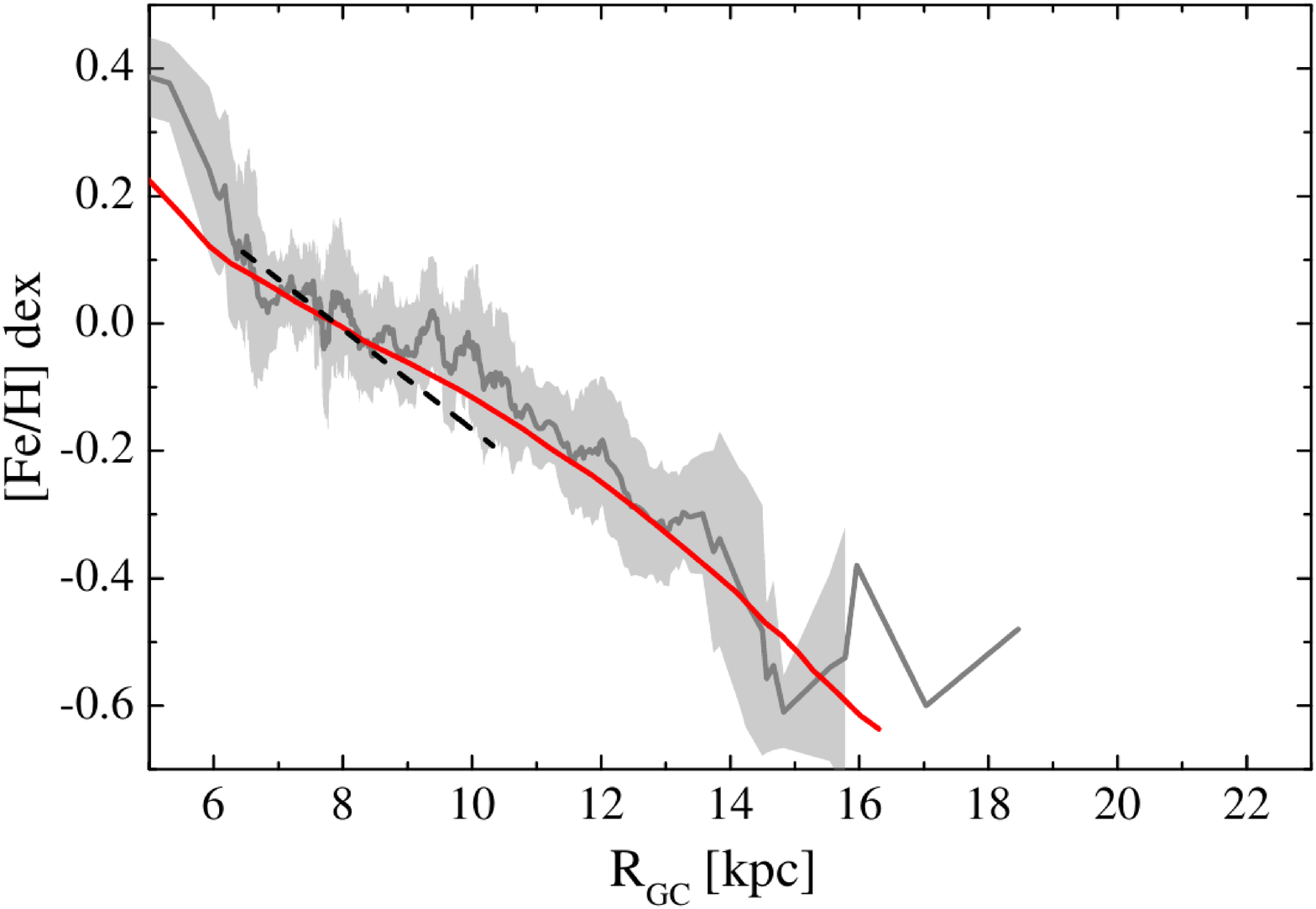}
\includegraphics[scale=.14]{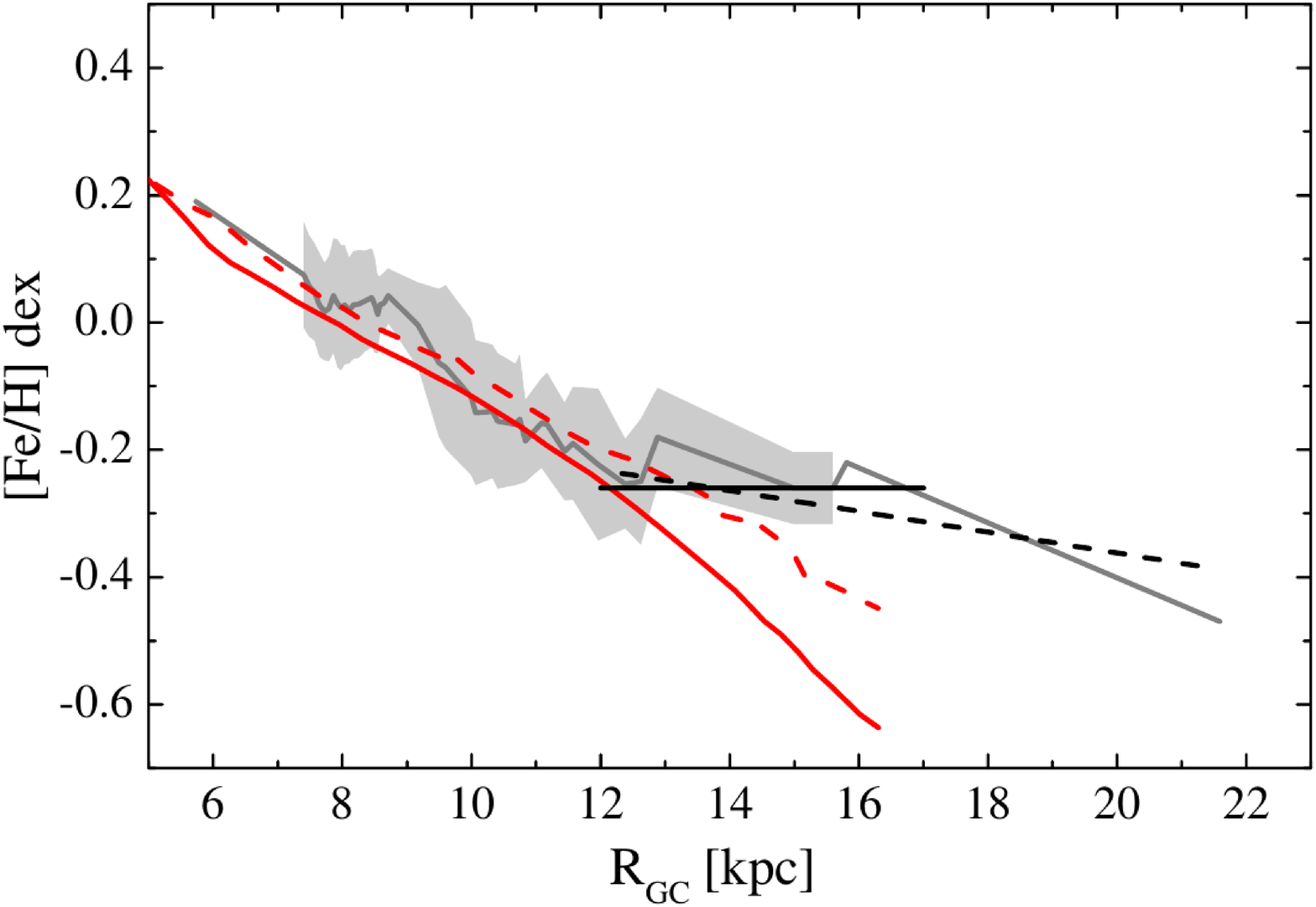}
\caption{Radial metallicity gradient derived from Cepheids and open cluster data compiled, respectively, by \citet{genovali2014} and \citet{netopil2016}. {\it Left} panel compares the youngest populations of these tracers with models. The grey line and the light grey area show the running average and the spread in iron of the Cepheid data. The black dashed line shows the gradient derived from a subset of young open clusters with ages less than 0.5\,Gyr. For comparison, the red line indicates the gradient for the gas from the Galactic chemical model by \citet{minchev2014}. The {\it right} panel presents the metallicity gradient derived from an intermediate population. The grey area shows the running average obtained from open clusters with ages between 0.5 and 3.0\,Gyr. The horizontal black solid line is the mean value of those clusters located in the outer disk, with galactocentric distances larger than 12\,kpc. The black dashed line is the gradient derived for all clusters in the outer disk. For comparison, the red dashed line shows the Galactic chemical model by \citet{minchev2013} that includes the dynamical effects of the Galactic disk after 2\,Gyr. This Figure is taken from \citet{netopil2016}, see this paper for details}
\label{fig:oc}  
\end{figure}

Open clusters are, apart from Cepheids (Sect.~\ref{chemo-cep}), the best tracers to derivate  metallicity  over large ranges in Galactocentric distance. However, accurate and homogeneous observations of open clusters are required to properly trace the chemical abundance distribution in the Galactic disk. Recently, \citet{jacobson2016} showed that the GES open clusters at galactocentric distances $[5.5, 8]$\,kpc exhibit a [Fe/H] radial metallicity gradient of $-0.10\pm0.02$\,dex/kpc consistent with values obtained by \citet{netopil2016} with a larger open cluster sample (172 open clusters) at distances R=[6,14]\,kpc compiled using different criteria. Both data sets are showing us no evidence for a steepening of the inner disk metallicity gradient inside the Solar circle. Furthermore, \citet{netopil2016} compared open clusters and Cepheids with the
Galactic  chemical model by \citet{minchev2013} (see Fig.~\ref{fig:oc}). Open clusters show a much flatter gradient than Cepheids or these model predictions. Possible explanations proposed for this flattening are an underestimation of radial migration in the models or other not yet included  mechanisms acting at large galactocentric radius. It is evident that more open cluster data is needed in the outer Galactic disk. Open clusters are introducing an important constraint to the chemical evolutionary models and its time evolution scenarios. Note that other independent data, in this case from APOGEE DR10 and with high-resolution metallicity measurements, reproduce interesting features such as the lack of gradient in the logarithmic $\alpha$-elements-to-metal ratio, $[\alpha$/M],  across the range $7.9<R< 14.5$\,kpc (\citealt{frinchaboy2013}).

Concerning field stars, the GES FGK dwarf and giant survey allows, for the first time, to extend  high-resolution spectroscopic surveys for field stars further away from the Solar vicinity. \citet{mikolaitis2014} published a first radial metallicity abundance gradient of eight elements (Mg, Al, Si, Ca, Ti, Fe, Cr, Ni, and Y) from this survey. The bimodality observed in the [Mg/M] distribution was attributed to thick and thin disk populations. Both  radial (4 to 12\,kpc) and vertical (0 to 3.5\,kpc) gradients in metallicity were derived for the thin and the thick disk separately. The interpretation of this huge amount of data is complex, with the thin disk showing a positive radial $[\alpha$/M] gradient in contrast to the flat gradient derived for the thick disk population. Concerning vertical gradients, the thick disk hosts shallower a negative vertical metallicity gradient than the thin disk in the Solar cylinder. These and upcoming new analyses of GES, APOGEE and SEGUE data, among others,  are being  examined in the context of new and complex models: from detailed analytic local disk models (e.g., \citealt{just2015}) to $N$-body  simulations of Milky Way-like galaxies (e.g., \citealt{kawata2016}; \citealt{roca2016}). In all cases, kinematic data will play a fundamental role here (see, e.g.,  \citealt{kordopatis2016}).

\subsection{The ``Outside-in'' versus ``Inside-out'' Disk Formation Scenarios}
\label{chemo-disc}

The comparison of observations and models involve global properties of the Milky Way such as rotation curves, scale lengths, masses (gas, stars and dark matter), gas flows, and stellar abundances\index{disks, formation scenarios}. Moreover,  dynamical  aspects  also play a critical role. They can be analyzed and taken into account through analytical or semi-analytical models or from cosmological $N$-body plus hydrodynamical models, the latter ones needed to assess the effect of the environment and satellite accretion on shaping the outer disks. Several efforts have been conducted up to now in this direction. As an example of the heterogeneity in the obtained results we can mention the works of \citet{cescutti2007}, \citet{pilkington2012b}, \citet{haywood2013} and \citet{minchev2014}. Whereas the first three proposed  an inside-out disk formation  scenario, \citet{haywood2013} put forward an outside-in scenario to explain  Solar neighbourhood data. These models are all only partially compatible with GES data (Sect.~\ref{chemo-GES}), and others have only been tested against external galaxy data (e.g., \citealt{pilkington2012a}). The addition of  upcoming new observational constraints to these models is complex. Although a general prediction of inside-out formation would be a fast decrease in $[\alpha$/Fe] with increasing radius for stars in a narrow metallicity range, different slopes for the thin and thick disk population (GES results, see \citealt{mikolaitis2014}) increase the complexity. This research field will be a key priority for the next decade. It will be possible to combine {\it Gaia} and spectroscopic data from the Milky Way with extragalactic work such as that done in the context of the CALIFA or MaNGA surveys.

\section{Large Surveys in the Next Decade}
\label{LargeSurvey}

The beginning of the 21st century is, without doubt, a new golden age for  Galactic astronomy. Data coming  from both space missions and ground-based surveys would be a dream in 1962, when Eggen, Lynden-Bell and Sandage derived the first model of galaxy formation and evolution (\citealt{ELS1962}).  Astrometric ({\it Gaia}, the Large Synoptic Survey Telescope: LSST) and spectroscopic ground-based surveys are fully complementary, providing both local and global measures of the six-dimensional phase-space distribution function, needed to encode valuable dynamical information. The status and products of the {\it Gaia} mission (2014-2022) are detailed in  Sect~\ref{gaia}. In the coming years, more than $10^{9}$ stars up to {\it Gaia} apparent magnitude $G\sim20$ will have an astrometric accuracy never envisaged up to now. These data will be complemented with the data collected by the LSST. This telescope, expected to be fully operational in 2022,  will measure proper motions with an accuracy of about 1 mas/year to a limit 4 magnitude fainter than the {\it Gaia} survey. Geometric parallaxes will  also be provided by LSST, with an accuracy similar to that of the faintest stars in {\it Gaia} ($\sim 0.3$\,mas)  at $V=20$, and up to $\sim$3\,mas at $V=24$. The acquisition of good signal-to-noise spectra at intermediate and high resolution is mandatory to complement these accurate astrometric data with good  radial velocities and chemical abundances for a significant number of stars. In Table~\ref{tab:1} we present a non-exhaustive list of the most promising spectroscopic surveys ongoing or planned for the near future. A detailed description of the capabilities of each of these surveys can be found in their respective web pages. Here, we will focus on one of them, the WEAVE\footnote{See http://www.ing.iac.es/confluence/display/WEAV/The+WEAVE+Project} spectroscopic survey. This survey, to be undertaken in the Northern hemisphere, will open us a new window to the Galactic anticentre, and thus to the  outskirts of the Milky Way.

WEAVE is a multiobject spectroscopic instrument using about one thousand fibers that will be operated at the William Hershel telescope (WHT, Canary Islands). Its science case and survey plan are almost fixed. During the first five years of operation (2018-2022), several Galactic plane radial velocity surveys will be executed using as kinematic tracers  disk Red Clump and young OBA-type stars. WEAVE's low-resolution mode ($R=5000$) will provide radial velocities for millions of stars up to the {\it Gaia} limiting magnitude. We estimated to obtain radial velocities for about one million Red Clump stars with an accuracy better than $\sim 5$\,km/s. Expected accuracies for stars brighter than $V=18$ are even better, with errors as small as $1-3$\,km/s. Complementarily, several chemical tagging programs for Galactic archeology will be undertaken using the high-resolution mode of WEAVE ($R=20000$).   Both kinds of data are critical ingredients to establish the chemical gradients of the Galactic thin and thick disk (Sect.~\ref{chemo-grad}) and the dynamics of the non-axisymmetric components of the outer regions of the MW (Sect.~\ref{nonaxi}).  A promising survey of well-selected open clusters (ages $>$100 Myr) will significantly contribute to the derivation of the radial metallicity gradient from these excellent tracers in the outer regions of the Milky Way. 

\subsection{The {\it Gaia} Mission}
\label{gaia}

The high-quality astrometric and photometric data already collected by {\it Gaia}\index{{\it Gaia}} during its first almost two years of successful scientific operation allow us to anticipate that, in few years, {\it Gaia} will revolutionize our understanding of our own Galaxy, the Milky Way, and its surroundings. In this Section we will briefly describe the {\it Gaia} products to be published in the first and end-of-mission data releases (end of 2022). More in detail, we will quantify the expected accuracy in position, transversal and radial velocity, and stellar astrophysical parameters of some the key stellar tracers of the outer  regions of the Milky Way. 


\runinhead{{\it Gaia} products.} %
The {\it Gaia} data release scenario, based on the current status of data processing, is described in the {\it Gaia} web page\footnote{{\it Gaia} Science Performance web page:  http://www.cosmos.esa.int/web/gaia/science-performance}. The first {\it Gaia} data release ({\it Gaia} DR1) has been made public on September 14th, 2016. This release, ready for the first {\it Gaia} science exploitation tasks,  has provided to the open community the first 3D map of the Solar neighbourhood, with parallaxes and proper motions with unprecedented accuracy for more than two million Tycho sources. Estimated astrometric errors are  discussed in detail by \citet{michalik2015} and validation tasks indicate that these expectations are being accomplished. The accuracy in astrometric data for Hipparcos sources ($\sim 10^{5}$ stars) will improve by a significant factor. Parallax accuracy will be improved by a factor $2-10$, and proper motions by a factor more than  $25-30$. For  Tycho sources  ($\sim 2 \times 10^{6}$ stars) we will have an astrometric accuracy as good as that published for the Hipparcos sources in 1997. It is not until the Second {\it Gaia} data release ({\it Gaia} DR2) that the 22 months of mission operation will provide the five astrometric parameters (position, parallax and proper motions) for all sources up to $G\sim 20$ (only single-star behaviour). As described below, with this accuracy we will have good distances for some of our stellar tracers (see Sect.~\ref{pop-res}) up to $3-4$\,kpc from the position?of the Sun. Also in this release, expected for end 2017, integrated photometry from the  BP and RP instruments, basic stellar astrophysical parameters  and mean radial velocities for bright sources will be published. The publication of  BP/RP spectra (\citealt{jordi2010}) for well-behaved objects,  orbital solutions for binaries and  object classification and astrophysical parameters will need to wait until 2018-2019. The final release is expected for 2022.  The  {\it Gaia} Archive core system web interface \footnote{{\it Gaia} Archive web page: http://gaia.esac.esa.int/archive/ } has been open from the date of the first release. Its continuous updates, both for tools and data, will be a reference for  upcoming studies. In the following, we present some examples of the {\it Gaia} data capabilities. They have been derived using the fortran code first used in \citet{romero2015} \footnote{The code was released at the 2nd {\it Gaia} challenge workshop and is publicly available at \texttt{https://github.com/mromerog/Gaia-errors}}.

\begin{figure}[t]
\centering
\includegraphics[width=\textwidth]{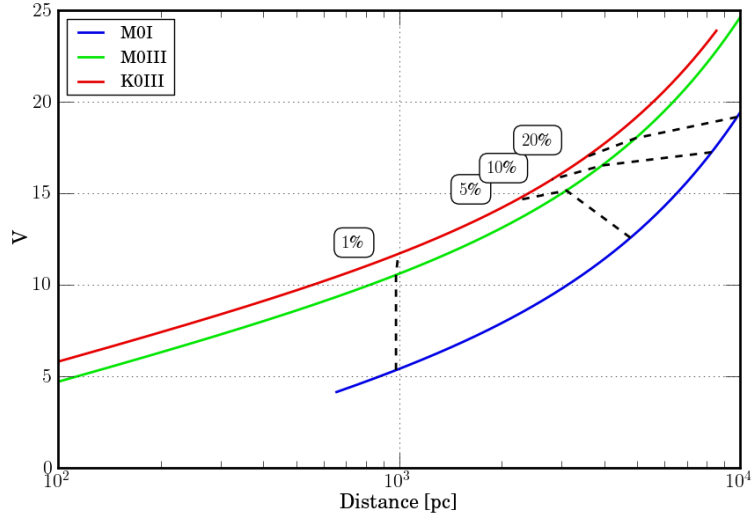}
\caption{Mean relative parallax accuracy horizons for red giants and supergiants. The
plot of visual apparent magnitude versus heliocentric distance has been made  assuming an extinction of 1\,mag/kpc, a good approximation for the  interstellar absorption toward the Galactic anticentre . Dashed lines represent the constant  mean relative parallax accuracy expected for {\it Gaia} DR2 expected for the end of 2017 (code courtesy of A. Brown)}
\label{fig:2}  
\end{figure}

\runinhead{Parallax horizon accuracy (tracers).}  
In Fig.~\ref{fig:2} we show the accuracy expected for the {\it Gaia} trigonometric parallax data to be published in the second release (end 2017). We can see how Red Clump stars placed as far as $3-4$\,kpc toward the Galactic anticentre will have distance accuracies better than  20\%. As discussed before (see Sect.~\ref{pop-res}), Red Clump  giant and supergiant stars are good trace populations  for both Galactic structure and kinematic studies.

\begin{figure}[t]
\centering
\includegraphics[scale=.32]{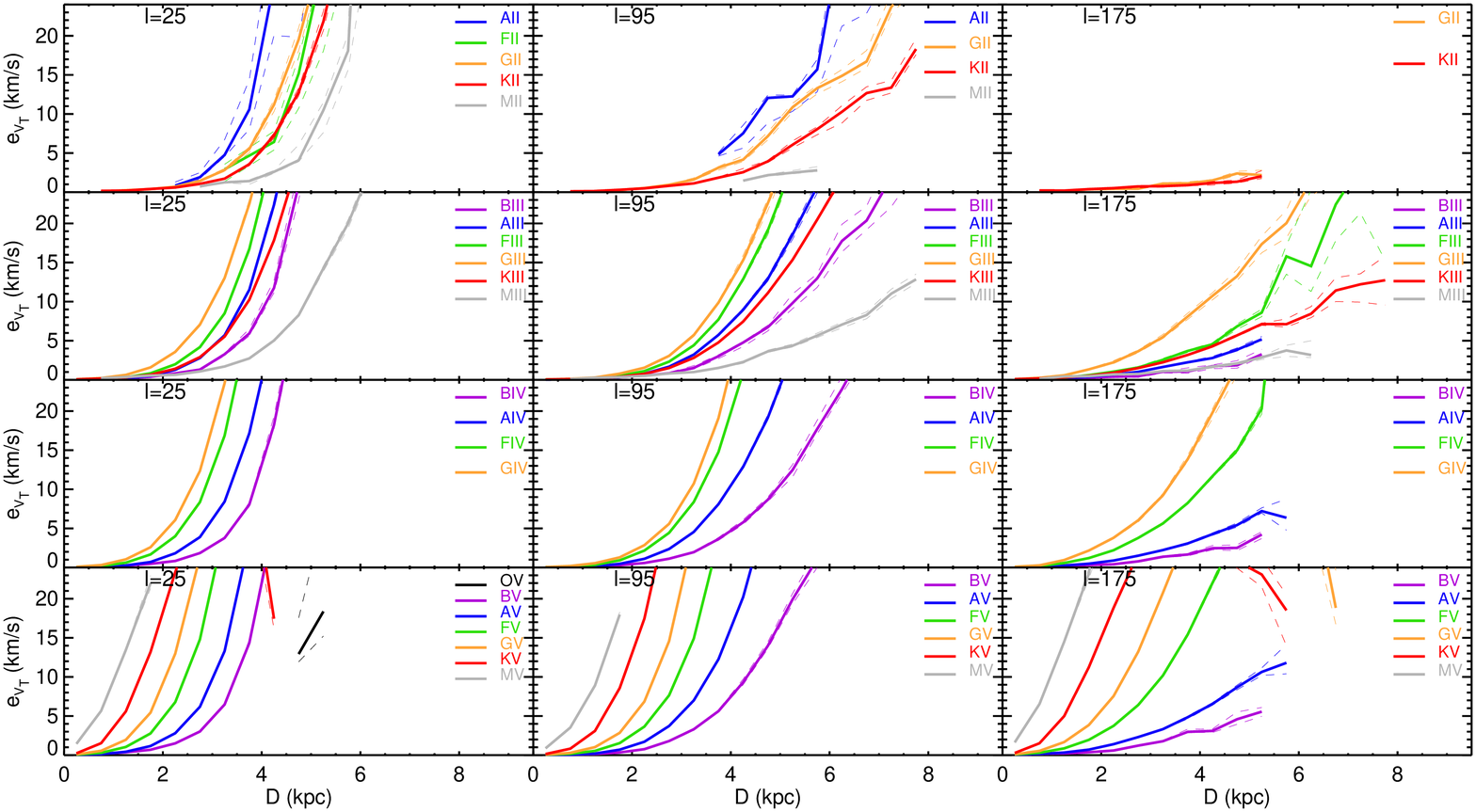}
\caption{Median error in transverse velocity ($V_{\rm T}$) as a function of heliocentric distance expected from end-off-mission {\it Gaia} data. The stellar populations shown are those of the {\it Gaia} Universe Model Simulator (\citealt{robin2012}). Three different directions in the Galactic plane have been considered. The dashed lines show the 75\% confidence limit of the median. The locations of the main resonances CR, ILR 2:1, ILR 4:1, OLR 4:1, OLR 2:1 of the tight-winding approximation spiral arms are shown with black horizontal lines (solid, dashed, dotted, dashed-dotted, long-dashed, respectively). The rotation of the Galaxy is towards the left (courtesy of T.~Antoja)}
\label{fig:3}  
\end{figure}

\runinhead{Transversal velocity accuracy.}  

\citet{antoja2016} analyzed the {\it Gaia} capabilities to characterize the Milky Way spiral arms. In Fig.~\ref{fig:3} we reproduce the expected accuracies in the {\it Gaia} end-off-mission tangential velocities for several disk populations. As discussed by \citet{antoja2016}, errors in the median transverse velocity shall be $<1$\,km/s to distinguish spiral features and, furthermore, errors larger than 20\% could introduce biases in the spiral pattern characterization. A careful statistical treatment of {\it Gaia} data will be indispensable and, as much as possible, it is recommended to work directly in the space of the observables (see \citealt{romero2006}; \citealt{abedi2014}).

\runinhead{Radial velocities and chemistry.} 

The {\it Gaia} spectro-photometers BP and RP (\citealt{jordi2010}) provide  low-resolution spectra, with resolutions $R= 40-25$ and  $R=130-70$ in the blue and red bands, respectively. Being conservative, one can expect to derive end-off-mission astrophysical parameters with an accuracy of about $75-250$\,K in effective temperature, $0.2-0.5$\,dex in $\log(g)$, $0.1-0.3$\,dex in [Fe/H] and $0.06-0.15$\,mag in $A_{V}$ for a star with {\it Gaia} magnitude $G\sim15$ (\citealt{bailer2013}). From the {\it Gaia} intermediate resolution spectrograph ($R\sim 11000$) is expected to yield  radial velocities for Red Clump stars with $V<15$, excellent tracers on the Milky Way outskirts (see Sect.~\ref{pop-res}), with an accuracy of $\sim5$\,km/s. We estimate {\it Gaia} will provide us with  about $10^{6}$ of such objects. The expected radial velocity error will increase  to  $\sim10$\,km/s for the $2\times10^{6}$ Red Clump stars to be observed by {\it Gaia} with $15 < V < 16$  (see \citealt{romero2015} for details). 


\begin{figure}[t]
\centering
\includegraphics[width=0.7\textwidth]{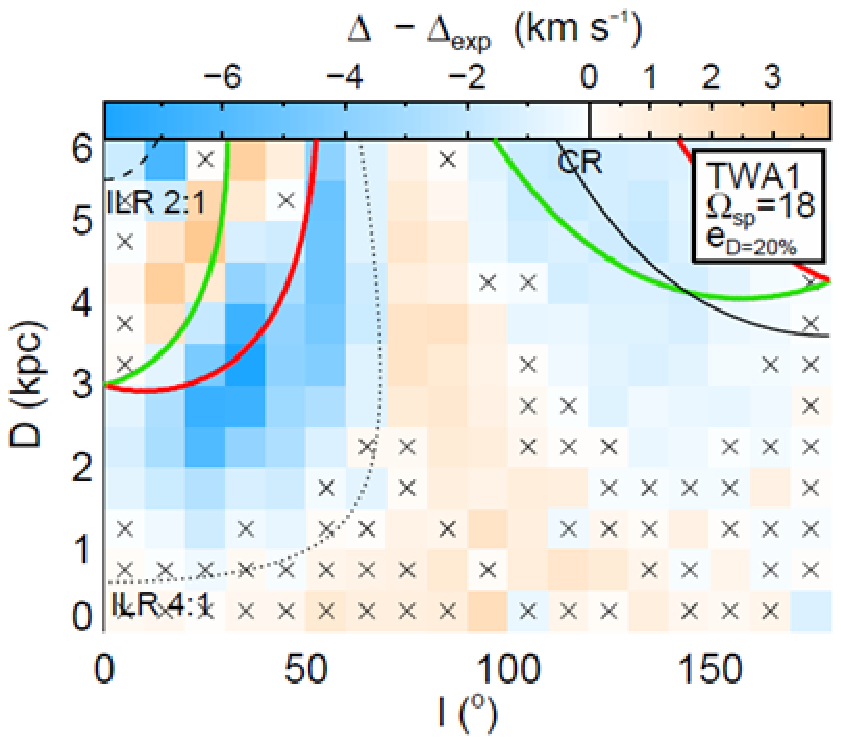}
\caption{Residuals in the tangential velocities expected from a simulated realistic density wave spiral arm perturbation. See text and \citet{antoja2016} for a detailed explanation (courtesy of T. Antoja)}
\label{fig:5}   
\end{figure}

\runinhead{New insights into spiral arm nature.}
Up to now, most studies have been done considering only radial velocity data, proper motions at large distances being too uncertain to be used. Recently, \citet{antoja2016} have proposed a new independent method to disentangle the nature of the spiral arm\index{spiral arms, nature} using only  astrometric data from {\it Gaia}. This method is based on the comparison of stellar kinematics of symmetric Galactic longitude regions in the Galactic plane. 
In Fig.~\ref{fig:5} we show the differences in tangential velocities expected when comparing two regions symmetric in Galactic longitude. In this case  a density wave spiral arm model has been considered, and the expected contribution of the axisymmetric galactocentric motion has been subtracted.  It is important to note that the analysis is done using only {\it Gaia} data. This strategy allow to more clearly identify possible biases introduced when propagating errors or when combining variables. As can be seen, typical kinematic trends induced only by the spiral arms are of the order  of $\sim2-10$\,km/s in tangential velocity. As expected,  these differences are very small in the  Galactic anticentre telling us that it will be hard to directly derive the kinematic perturbations at large distances using only {\it Gaia} data. Fortunately, {\it Gaia} astrometric data will be combined with spectroscopic data  from ground-based surveys (see Sect.~\ref{LargeSurvey}). Nonetheless, the precision on median transverse velocity obtained exclusively with {\it Gaia} data are $\sim1$\,km/s up to $\sim4-6$\,kpc for some giants stars, and $\sim0.5$\,km/s up to $\sim 2-4$\,kpc for subgiants and dwarfs (\citealt{antoja2016}).

\begin{figure}[t]
\sidecaption[t]
\includegraphics[scale=.36]{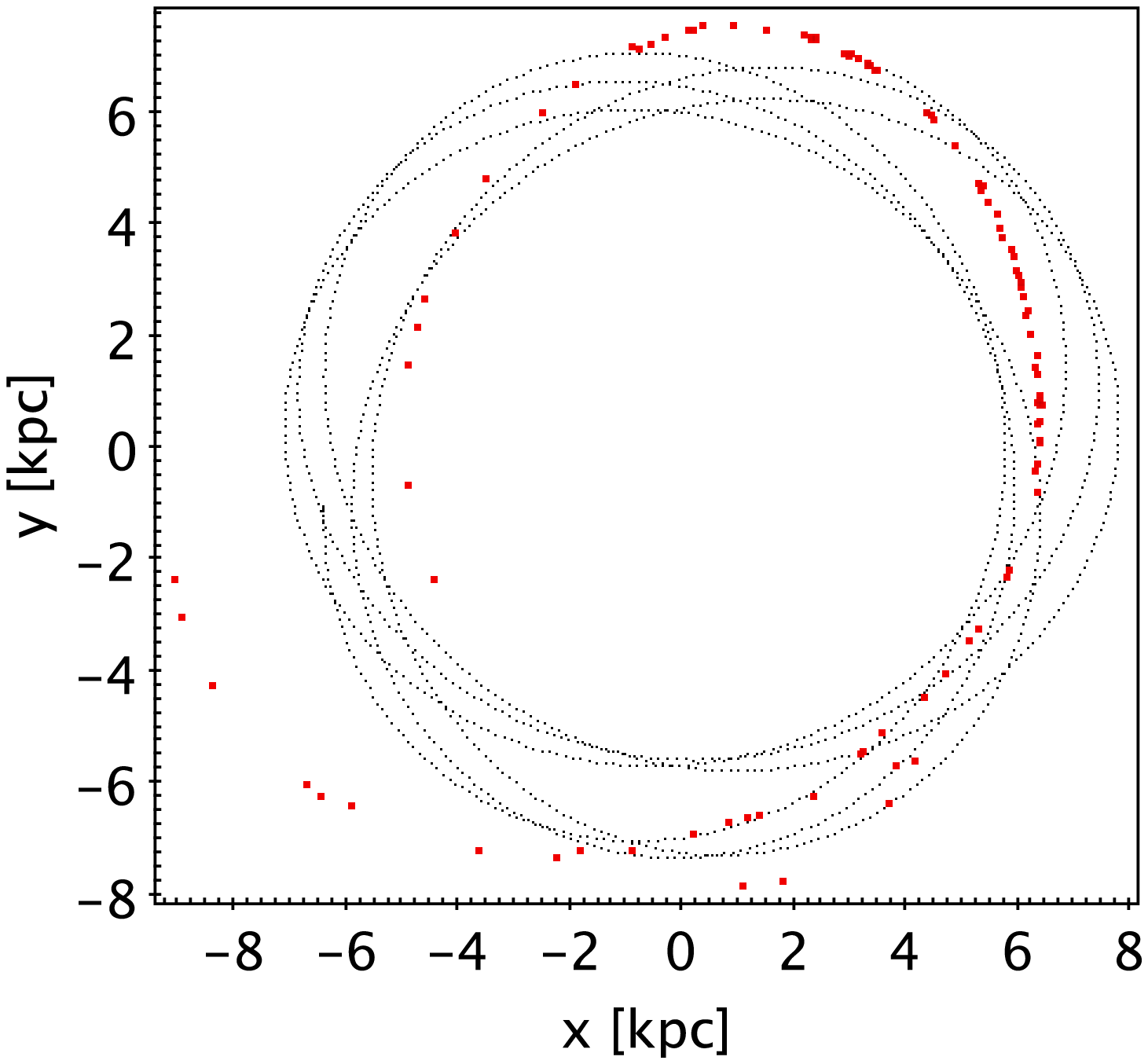}
\includegraphics[scale=.36]{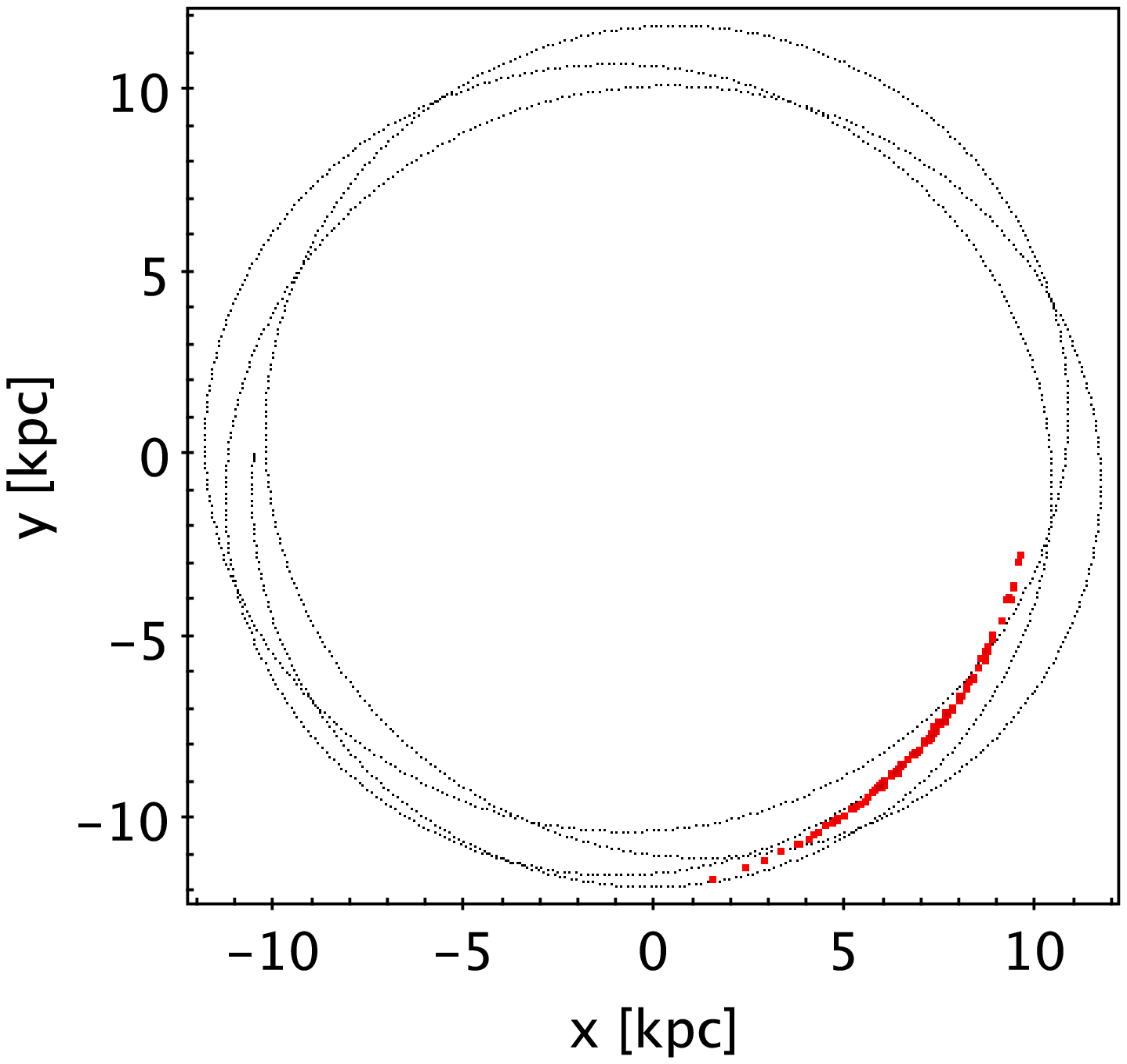}
\caption{Quantitative evaluation of the {\it Gaia}-DR2  plus WEAVE capabilities to  derive the  birth positions of two hypothetical Red Clump stars, one placed near the Perseus arm toward the anticentre ({\it left}) and the other near the Sagitarius arm ({\it right}). See text for a detailed explanation}
\label{fig:4}   
\end{figure}

\runinhead{Orbital analysis back in time.}
In Fig.~\ref{fig:4} we show the orbital trajectory back\index{orbital motions, Galactic} in time of two simulated test particles. On the left we simulated a Red Clump star placed near the Perseus arm and towards the Galactic anticentre as seen from the  position of the Sun (galactocentric coordinates $(X,Y) = (-10.5, 0)$\,kpc). Its apparent magnitude would be $V\sim 15$. The second particle, in the right figure, simulates the same target but placed inside the Sun's galactocentric radius, with coordinates $(X,Y) = (-6.5, -1.0)$\,kpc, that is, near the Sagitarius arm, with an apparent magnitude $V\sim 16$. One hundred realizations of each particle have been done assuming a Gaussian distribution of the astrometric proper motions errors  as expected from {\it Gaia} DR2 and a radial velocity error as expected from the WEAVE spectroscopic survey ($\sigma_{V_{\rm r}} \sim 1$\,km/s). Orbital evolution back in time has been performed using a realistic Galactic potential (\citealt{romero2015}). In black, we show the trajectories of these particles and as points (in red) the final positions of these one hundred realizations after evolving   $\Delta t = 1$\,Gyr on the Galactic potential. The dispersion of these points gives an indication of the accuracy we will have when looking, for example, at the  birth position of a particle in the Perseus or in the Sagitarius arm. We have confirmed that the WEAVE contribution to improve the radial velocity component is crucial for this detailed analysis of the  birth positions of field stars or open clusters.



\section{Conclusions}

The high-quality astrometric and photometric data already collected by {\it Gaia} during its almost two years of successful scientific operation to date allow us to anticipate that, in few years, {\it Gaia} and the upcoming ground-based spectroscopic surveys will revolutionize our understanding of the Milky Way. We have shown the capabilities we will have to characterize the kinematic and chemical properties of the resolved stellar populations in the outer regions of the Galactic disk. A detailed and robust statistical treatment of all these data will provide us with a local and global measure of the six-dimensional phase-space distribution function, essential to encode valuable dynamical information, whereas the spectroscopic surveys  will provide us with the chemical characterization of the stellar populations. These are the ingredients needed to constrain new and powerful chemodynamical models for the Milky Way. Novel approaches in the theoretical and modelling work are required to treat this huge and unprecedentedly accurate amount of data. It is mandatory to use both bottom-up and top-down dynamical approaches. Furthermore, the {\it Gaia} archive, designed to maximize the scientific return from the {\it Gaia} mission, will be a dynamic platform to implement new and powerful software which would support scientific research in this field. These efforts and the huge improvements in the extragalactic domain (e.g., by the CALIFA or MaNGA surveys) will definitively provide new insights on the formation and evolution of galactic disks.

\begin{acknowledgement}
This work was supported by the MICINN (Spanish Ministry of Science and Innovation) - FEDER through grant AYA2012-39551-C02-01 and ESP2013-48318-C2-1-R, and by the European Community's Seventh Framework Programme (FP7/2007-2013) under grant agreement GENIUS, FP7-606740. I thank my colleagues at  Barcelona University, and specially Merc\`{e} Romero-G\'{o}mez, Teresa Antoja, Maria Mongui\'{o}, Santi Roca-F\`{a}brega and Roger Mor for our common research activities  in this exciting {\it Gaia} Era. 
\end{acknowledgement}

\bibliographystyle{spbasic}
\bibliography{ref}

\printindex

\end{document}